\def\dk{{\rm k}} 
\def\henon{H\'{e}non} 
\def\MG{M_{\rm G}}
\def\nbodysix{{\small\texttt{NBODY6}}}
\def\nbodyfour{{\small\texttt{NBODY4}}}
\def\emacss{{\small\texttt{EMACSS}}}
\def\tdiss{\tau_{\rm diss}} 
\def\trh{\tau_{\rm rh}} 
\def\tcr{\tau_{\rm cr}}

\def\mmean{\bar{m}}
\def\msun{M_\odot}

\def\rh{r_{\rm h}}
\def\rv{r_{\rm v}}

\def\rj{r_{\rm J}}

\def\pc{{\rm pc}}

\def\Ra{R_{\rm a}}
\def\RG{R_{\rm G}}
\def\Rp{R_{\rm p}}

\documentclass[useAMS, usenatbib]{mn2e}
\usepackage{amssymb,latexsym,graphicx,natbib,eufrak,times,amsmath,xfrac,nicefrac}
\usepackage{CJKutf8}

\title[Evolution of star clusters on eccentric orbits]{Evolution of star clusters on eccentric orbits}
\author[Cai et al.]
 {Maxwell Xu Cai (\begin{CJK*}{UTF8}{gbsn}蔡栩\end{CJK*})$^{1,2}$\thanks{Also known as: Maxwell Tsai. E-mail:
maxwell@nao.cas.cn} Mark Gieles$^{3}$, Douglas C. Heggie$^{4}$ and Anna Lisa Varri$^{4}$\\
$^{1}$National Astronomical Observatories of China, Chinese Academy of Sciences, 20A Datun Rd., Chaoyang District, 100012, Beijing, P.R. China\\
$^{2}$Kavli Institute for Astronomy and Astrophysics, Peking University, Yi He Yuan Lu 5, Haidian District, Beijing 100871, P.R. China\\
$^{3}$Department of Physics, University of Surrey, Guildford GU2 7XH, UK\\
$^{4}$School of Mathematics and Maxwell Institute for Mathematical Sciences, University of Edinburgh, Kings Buildings, Edinburgh, UK-EH9-3JZ
}
\begin{document}
\date{Accepted 2015 October 5.  Received 2015 September 29; in original form 2015 August 3}


\pagerange{\pageref{firstpage}--\pageref{lastpage}} \pubyear{2015}
\def\LaTeX{L\kern-.36em\raise.3ex\hbox{a}\kern-.15em
    T\kern-.1667em\lower.7ex\hbox{E}\kern-.125emX}

\maketitle

\label{firstpage}


\begin{abstract}
We study the evolution of star clusters on circular and eccentric orbits using direct $N$-body simulations.
We model clusters with  initially $N=8\dk$ and $N=16\dk$ single stars of the same mass,  orbiting around a point-mass galaxy. For each orbital eccentricity that we consider, we find the apogalactic radius at which the cluster has the same lifetime as the cluster with the same $N$ on a circular orbit. We show that then, the evolution of bound particle number and half-mass radius is approximately independent of eccentricity. Secondly, when we scale our results to orbits with the same semi-major axis, we find that the lifetimes are, to first order, independent of eccentricity. When the results of Baumgardt and Makino for a singular isothermal halo are scaled in the same way, the lifetime is again independent of eccentricity to first order, suggesting that this result is independent of the Galactic mass profile.  From both sets of simulations we empirically derive the higher order dependence of the lifetime on eccentricity. Our results serve as benchmark for theoretical studies of the escape rate from clusters on eccentric orbits. Finally, our results can be useful for generative models for cold streams and cluster evolution models that are confined to spherical symmetry and/or time-independent tides, such as  Fokker-Planck models, Monte Carlo models, and (fast) semi-analytic models.
\end{abstract}

\section{Introduction}
The evolution of star clusters is driven by internal factors, such as two-body relaxation, stellar and binary evolution, and external factors such as the Galactic tidal field \citep[see e.g.][]{2003gmbp.book.....H}. As a result, star clusters gradually dissolve and eventually lose all their stars to the Galactic field. 

The escape rate from clusters in a static tidal field, as applies to cluster on circular orbits in a time-independent external potential, has been topic of extensive theoretical \citep*[e.g.][]{1961AnAp...24..369H, 1966AJ.....71...64K, 2011MNRAS.413.2509G} and numerical work \citep[e.g.][]{1990ApJ...351..121C, 1992ApJ...386..519O}. A consensus picture for the dependence of the dissolution time-scale $\tdiss$ on the properties of the cluster and its orbit has emerged. 

When approximating the tidal limitation by a simple cut-off radius, beyond which stars are considered unbound,  $\tdiss$ scales with the half-mass relaxation time-scale $\trh$ of the cluster, which itself depends on the number of stars in the cluster $N$, the crossing time of stars within the cluster $\tcr$   as $\trh\propto (N/\log\Lambda)\tcr$, where $\log\Lambda$ is the Coulomb logarithm, which slowly varies with $N$. For Roche-filling clusters on circular orbits in a scale-free Galactic potential, $\tcr$ is a constant fraction of the  period of the Galactic orbit, hence for these clusters $\tcr\propto\Omega^{-1}$ \citep{1987ApJ...322..123L}, with $\Omega$ the angular frequency of the orbit. 

If a tidal field is included, the escape of stars is delayed \citep{2000MNRAS.318..753F}, and this effect changes the $N$ dependence of $\tdiss$ to \citep{baumgardt01}

\begin{equation}
\tdiss(\RG, \epsilon=0)\propto \left(\frac{N}{\log\Lambda}\right)^{3/4}{\Omega^{-1}}.
\label{eq:tdisscirc}
\end{equation}
Here $\epsilon$ is the orbital eccentricity and $\RG$ is the Galactocentric radius, which for the circular orbit relates to $\Omega$ as $\Omega = V_{\rm circ}/R_{\rm G}$, with $V_{\rm circ}$ the circular velocity at $R_{\rm G}$.

For compact clusters that under-fill the Roche volume, the fraction of escapers per $\trh$ is lower because the tides are weaker, and because $\trh$ itself is shorter, the fraction of escapers per unit of physical time is approximately independent of the half-mass radius $\rh$ of the cluster and the result for $\tdiss$ of equation~(\ref{eq:tdisscirc}) is, therefore, almost independent of the initial $\rh$ \citep{2008MNRAS.389L..28G}.

\citet[][hereafter BM03]{bm03} studied clusters on circular and eccentric orbits using direct $N$-body integrations with \nbodyfour\ \citep{aarseth99}. Their clusters contained a stellar mass spectrum, the effect of stellar and binary evolution and a realistic description of the tidal field of the Galaxy, which was assumed to be due to a singular isothermal Galactic halo. For  eccentric orbits  they find that the $N$ dependence in $\tdiss$ is the same as for the circular orbits, and they show that $\tdiss$  can be expressed in terms of the scaling for the circular orbits as 

\begin{equation}
	\tdiss(\Ra=\RG, \epsilon) = \tdiss(\RG, 0) (1 - \epsilon).
	\label{eq:bm03}
\end{equation}
Here $\Ra$ is the apogalactic radius of the orbit, which in BM03 was kept the same as the Galactocentric radius $\RG$ of the circular orbit.

Along an eccentric orbit, the tidal field strength varies and it is, therefore, often assumed that the evolution of clusters on eccentric orbits is determined by the perigalactic radius $\Rp$, where the tidal field is strongest \citep*{1962AJ.....67..471K, 1983AJ.....88..338I}. This is indeed true for collisionless systems \citep*{2008ApJ...673..226P}, but it is not what follows from the BM03 result for collisional systems on eccentric orbits. Taken together, equations (\ref{eq:tdisscirc}) and (\ref{eq:bm03}) suggest that the ``effective radius'' $\RG^{\rm eff}$, i.e. the radius of the circular orbit on which a cluster has the same lifetime as a cluster on the given elliptic orbit, is given by $\RG^{\rm eff} = \Rp(1+\epsilon) = \Ra(1-\epsilon)$, i.e. the effective radius lies between $\Rp$ and $\Ra$. The different dependence of $\tdiss$ on the external tides as compared to the collisionless case, suggests that the combined influence of two-body relaxation and the (time-dependent) tides, result in a  different overall evolution of (globular) clusters than what is found for (collisionless) dwarf galaxies that get tidally stripped in the host potential \citep[see also the discussion in][on differences in the escape mechanisms in collisional and collisionless systems]{2015MNRAS.450..575A}.

In this study we want to establish whether it is possible to approximate the evolution of a cluster on an eccentric orbit, by that of a cluster on a circular orbit. Whether possible, or not, the answer  helps to identify the dominant mechanism that drives the escape from clusters on eccentric orbits. If possible, it would greatly simplify the treatment of eccentric orbits in dynamical models of cluster evolution that are limited to spherical symmetry/circular orbits and in (fast) semi-analytic models of clusters and cold tidal streams. Secondly, we aim to shed light on the scaling for $\tdiss(\Ra, \epsilon)$ for clusters on eccentric orbits. 

We run a series of direct $N$-body integrations of idealised systems, without the effect of stellar evolution, which can be scaled and compared to the result of BM03.
This paper is organised as follows: the details of the $N$-body experiments are described in Section \ref{sec:ic}. Our results are presented in Section \ref{sec:results} and our conclusions are presented in Section \ref{sec:conclusions}.

\section{$N$-body simulations}
\label{sec:ic}

\subsection{$N$-body integrator and units}
\label{ssec:nbody}
For all simulations we used the $N$-body code \nbodysix, which is a fourth order Hermite integrator with \citet{1973JCoPh..12..389A} neighbour scheme \citep{1992PASJ...44..141M, aarseth99, 2003gnbs.book.....A}, with accelerated  force calculation on NVIDIA Graphical Processing Units (GPUs) \citep{nitadori12}. All our models are scaled to the conventional \henon\ $N$-body units \citep{henon1971}, in which $G=M=-4E=1$, where $G$ is the gravitational constant and $M$ and $E$ are the total mass and energy of the cluster, respectively. Our models are initially in virial equilibrium, such that the gravitational energy $W=2E$ and the virial radius $\rv = -GM^2/(2W)=1$.

\subsection{Initial conditions}
\label{ssec:ics}
We model clusters with $N = 8\dk$ and $N = 16\dk$ point particles of the same mass, without primordial binaries, with initial positions and velocities sampled from a \citet{1911MNRAS..71..460P} model, truncated at ten scale radii\footnote{In principle, the Plummer model has no truncation radius; in practice, it is truncated at ten scale radii in {\tt NBODY6}.}. For this model $\rh\simeq0.78\rv$. 
The Galactic potential is that of a point mass and the differential
forces due to the Galaxy are added in a non-rotating frame that is initially
centred on the centre-of-mass of the cluster.

We adopt this simplified set of initial conditions because (i) we want to focus on one single physical ingredient (i.e. the tidal field), explored within the simplest possible choice of Galactic mass model, (ii) some of the key results of the paper are based on three different scaling of the simulations, which must therefore be performed in the absence of any factor imposing a physical scale (e.g. stellar evolution), (iii) we wish to provide some `empirical' evidence of the process underlying the escape of stars from clusters on elliptic orbits, for which a proper theory is still lacking, therefore we decided to explore first very idealised models, and to increase the complexity of the systems under consideration only in a second phase of the investigation.

For each $N$, we consider seven different orbital eccentricities: $\epsilon=[0.0, 0.1, 0.2, 0.3, 0.4, 0.6, 0.8]$. 
For the circular orbits, we choose an orbit such that  $\rh/\rj=0.1$, where $\rj$ is the Jacobi radius, which for the point-mass Galaxy and $\epsilon=0$ depends on $\RG$ and the mass of the Galaxy $\MG$ as
\begin{equation}
	\rj = \left(\frac{M}{3\MG} \right)^{1/3}\RG.
	\label{eq:rj}
\end{equation}

The initial conditions of \nbodysix\ need to be fed in physical units. We choose  $\MG = 10^{10}\,\msun$, $\rv=1\,\pc$ and $\mmean = M/N = 1\,\msun$. The remaining parameter to choose is $\RG$, which given the constraint of the initial $\rh/\rj$ and equation~(\ref{eq:rj}) is $\RG = 7.86(3\times10^{10}/N)^{1/3}\,\pc$, which is $\RG=1211\,\pc(962\,\pc)$  for the circular orbit of the $N=8\dk(16\dk)$ cluster\footnote{In this paper we use the definition of 1k = 1000, so that the 8k and 16k models correspond to the total particle number of exactly 8000 and 16000, respectively. Note that this is slightly different from the convention used in the BM03 paper, where they defined 1k = 1024.}. All models started with the same $\rh = 0.78$ in $N$-body units. The physical units are only used in the input of the code, and they are not relevant for our results and we report all our results in the internal $N$-body units (Section~\ref{ssec:nbody}).

We define $\tdiss$ as the time when 10\% of the initial number of stars remains bound. 
We then need to define bound. For a cluster on a circular orbit, in a coordinate system centred on the cluster and co-rotating with the Galactic orbit, bound is defined as having a Jacobi energy smaller than the critical energy of escape. For eccentric orbits, there is no conserved integral of motion, hence we need to find another way to separate bound from unbound stars. We consider a star as bound when the sum of its specific kinetic energy, computed from the velocities corrected for the centre-of-mass velocity, is less than its specific potential energy due to the $N - 1$ other stars, with $N$ being determined iteratively until convergence  \citep*[as in][]{renaud11}.

For each value of $\epsilon$ we aim to find the $\Ra$ that results in the same $\tdiss$ as for the circular orbit at $\RG$, i.e $\tdiss(\Ra, \epsilon) = \tdiss(\RG, 0)$. 
This is different from the approach of BM03, who started all their eccentric orbits at the same $\Ra(\epsilon)=\RG(\epsilon=0)$, which results in shorter lifetimes for the eccentric orbits. In Section~\ref{sec:results} we scale results for comparison. Because we do not know the scaling of $\Ra(\RG, \epsilon)$ \emph{a priori} for clusters with the same $\rh$, we find $\Ra$ by iteration: in a first attempt we adopt the scaling of BM03 (equation~\ref{eq:bm03}) and run a model with $\Ra = \RG/(1-\epsilon)^{2/3}$ (note that the index of $2/3$ is because for a point-mass Galaxy $\Omega \propto \RG^{-3/2}$). At this stage we could adapt the scaling $\tdiss(\Ra) \propto \Ra^{-3/2}$ to find $\Ra$ that results in the correct lifetime. 
However, scaling will not keep the initial half-mass radius fixed, which is our intention in this study, and we therefore proceed by finding the correct $\Ra$ by iteration. 

If $\tdiss$ of the first attempt is shorter(longer) than that of the circular orbits, we run an additional model with $\Ra$ 20\% larger(smaller).  
We continue this, until we have two models whose $\tdiss(\Ra, \epsilon)$ bracket the result for the circular orbit. We then apply a linear interpolation to get the final $\Ra$ and run a model at that $\Ra$. The final interpolated $\Ra$ values are summarised in Table~\ref{tab:data}. 

\begin{figure}
	\centering
	\includegraphics[width=8cm]{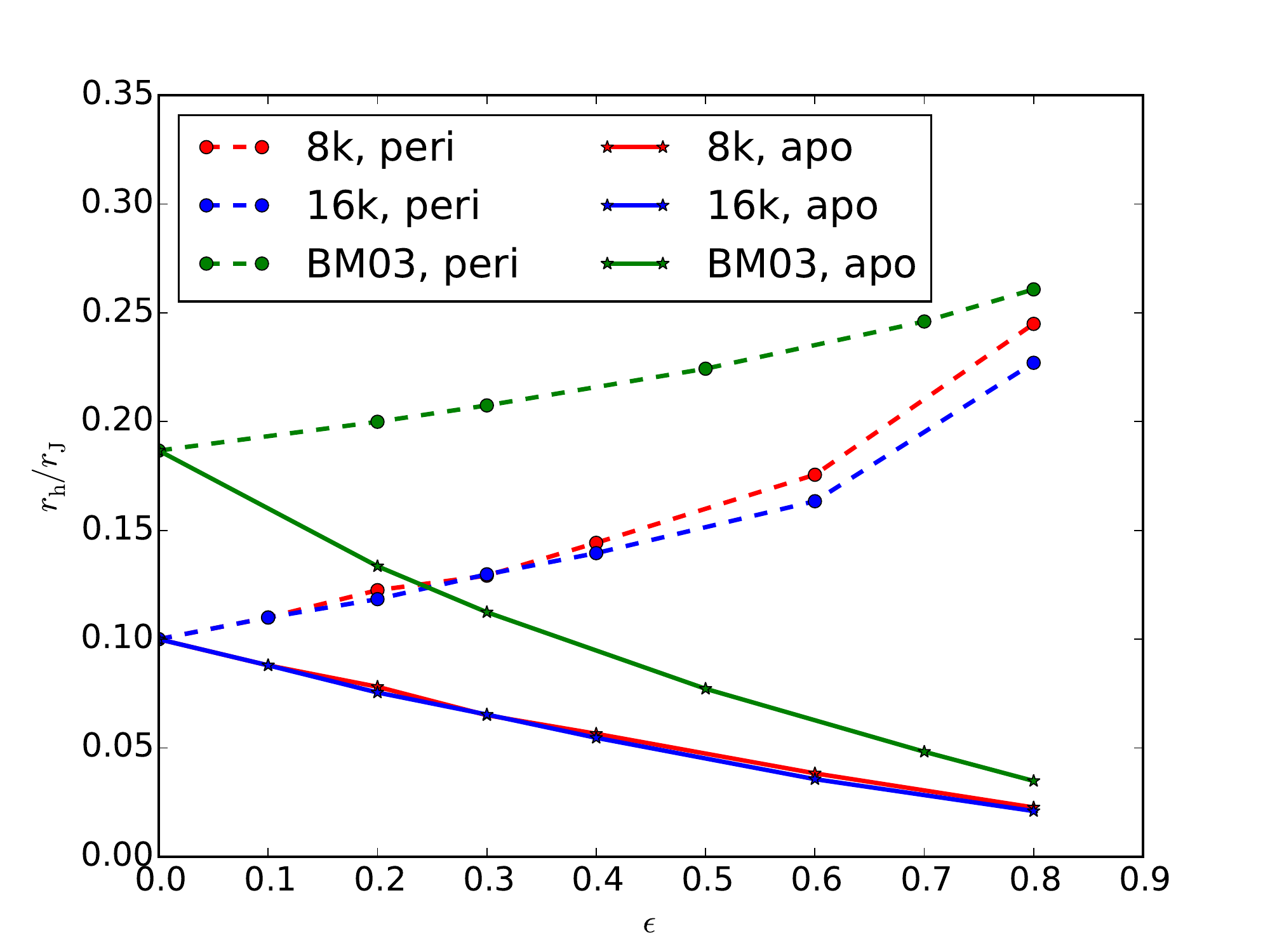} 
	\caption{Initial $\rh/\rj$ for all models as a function of orbital eccentricities $\epsilon$. the values corresponding to the apocentre are connected with full lines; the values corresponding to the pericentre are connected with dashed lines. Results for the $N=32 {\rm k}$ models of BM03 are also plotted in the same way for comparison.}
	\label{fig:rh_rj}
\end{figure}

The corresponding initial $\rh/\rj$ at apocentre for all models with different orbital eccentricities $\epsilon$ are shown in Figure~\ref{fig:rh_rj}, where $\rj$ was computed using equation~(7) in \citet{1962AJ.....67..471K}. For comparison, we present also the ratio $\rh/\rj$ the cluster would have if we would have started the evolution at pericentre. We also show the values for $\rh/\rj$ of the $N=32 {\rm k}$ models of BM03, which we will later compare our results against, using the \citet{1962AJ.....67..471K} definition for $\rj$ (note that the equation used by BM03 is slighly different).

 \begin{table} 
 \centering
\caption{Apogalactic radii $\Ra$ for the $N=8\dk$ and $N=16\dk$ simulations that result in the same $\tdiss$ as the circular model. The values of $\Ra$ for each $\epsilon$ were found by iteration, see the text in Section~\ref{ssec:ics} for details. }
  \label{tab:data}
  \begin{tabular}{llcccccccc}
\hline
   $N$       &        $\tdiss$                       &\multicolumn{7}{c}{$\Ra(\epsilon)$}\\
               &              & $0$  & $0.1$ & $0.2$ & $0.3$ & $0.4$ & $0.6$ & $0.8$  \\ \hline
  $8\dk$ &   5060        &1212 & 1362  &1516 &  1798  & 2043 & 2934 & 4834 \\
 $16\dk$&    8230        & 962 &  1081  &1245 &  1420  & 1677 & 2502 & 4126 \\
\hline
\end{tabular}
\end{table}

\section{Results}
\label{sec:results}

\subsection{Evolution of $N$ and $\rh$: can the evolution on an eccentric orbit be compared to the evolution on a circular orbit?}
\begin{figure}
	\centering
	\includegraphics[width=8cm]{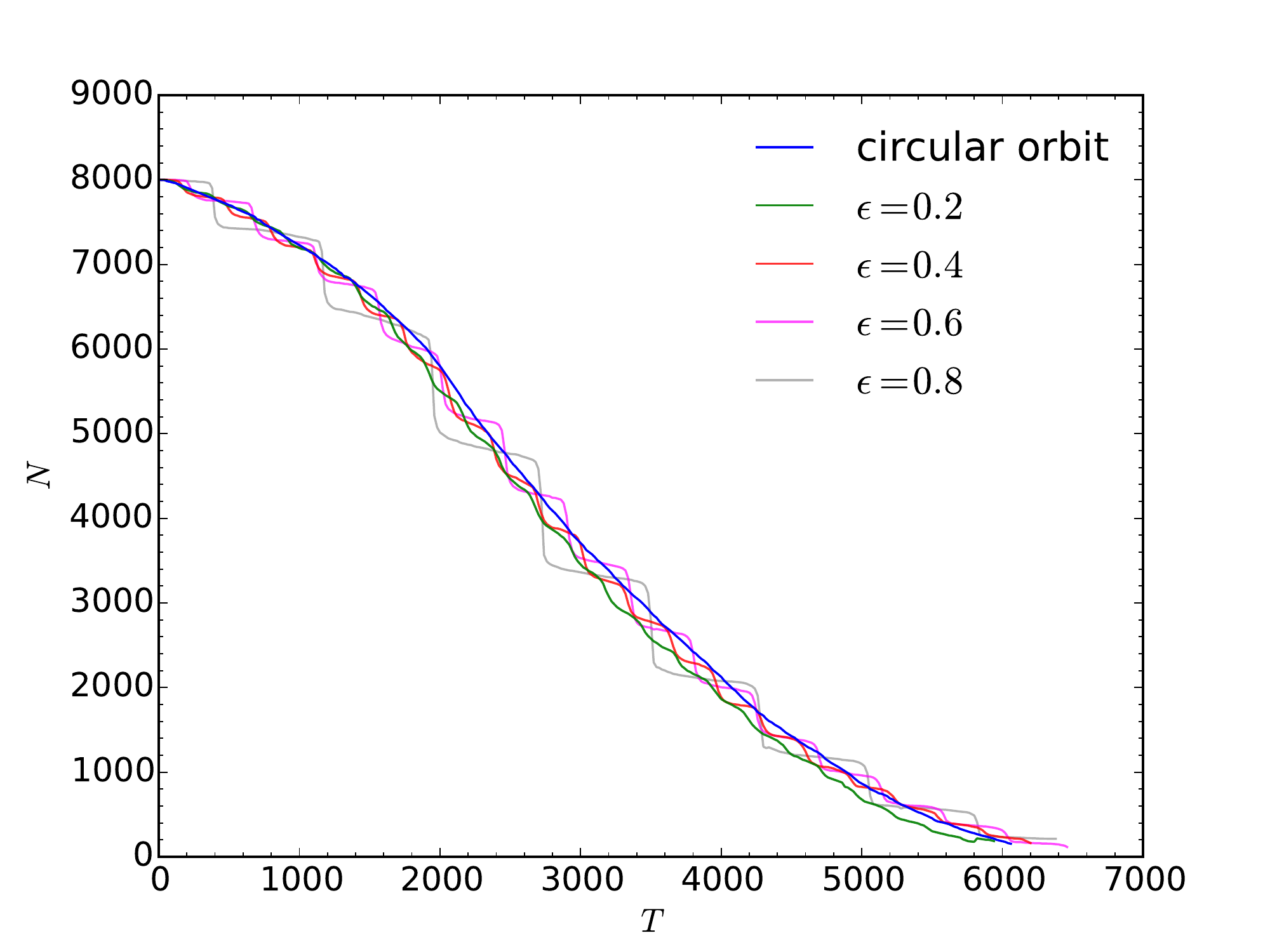} 
	\caption{Evolution of the number of bound stars for the $8\dk$ models with different orbital eccentricities $\epsilon$.}
	\label{fig:mbound_8k}
\end{figure}

\begin{figure}
	\centering
	\includegraphics[width=8cm]{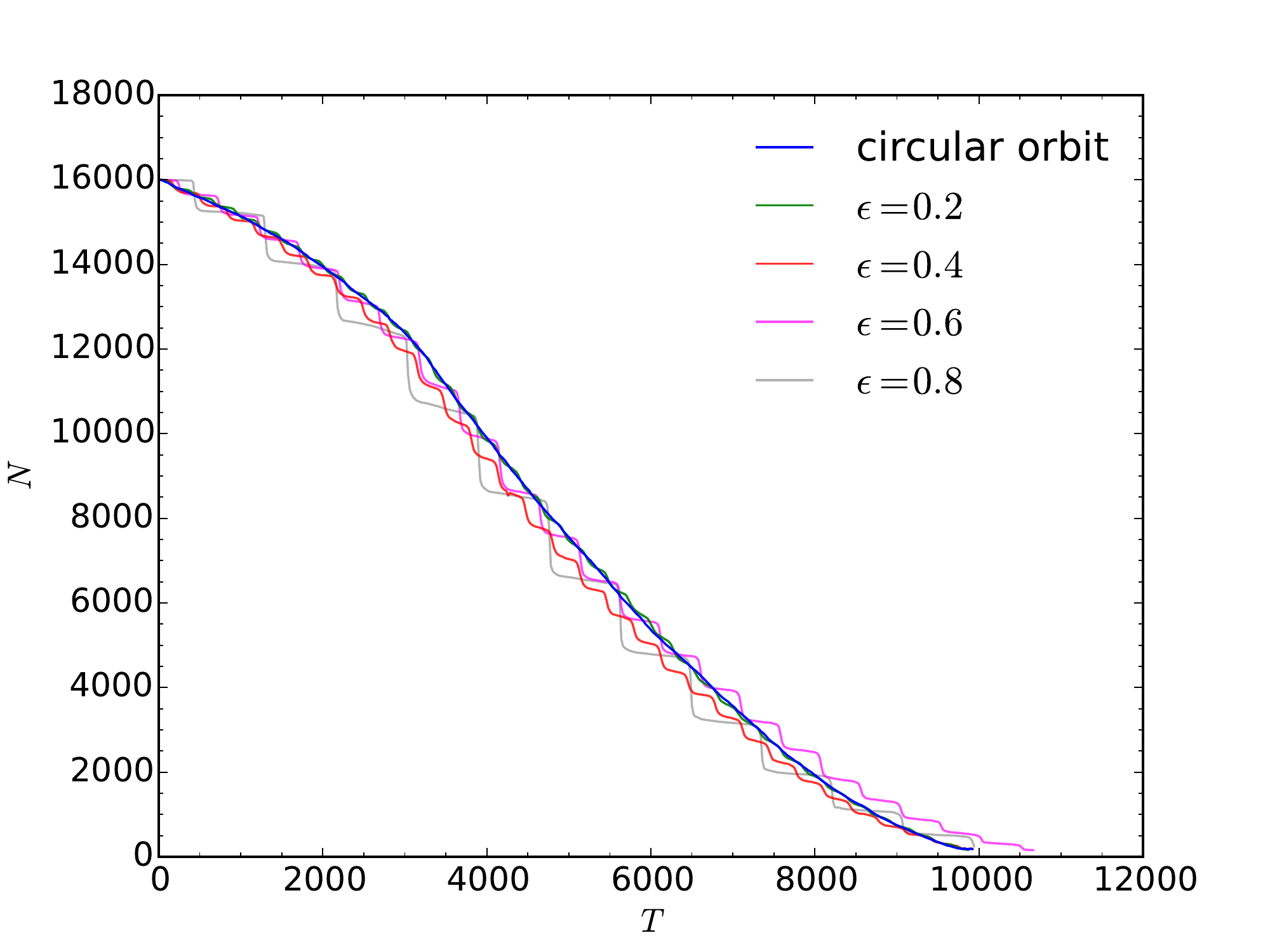} 
	\caption{Evolution of number of bound stars for the $16\dk$ models with different orbital eccentricities $\epsilon$.}
	\label{fig:mbound_16k}
\end{figure}

Figures \ref{fig:mbound_8k} and \ref{fig:mbound_16k} show the evolution of the bound $N$ for our models with different $\epsilon$ and initial $N=8\dk$ and $N=16\dk$, respectively. The $N(T)$ curves of the models on eccentric orbits display a `staircase' shape, with a frequency that corresponds to the orbital period. The amplitude of the `stairs' depends on the number of particles $N$ and the orbital eccentricity $\epsilon$. The steps correspond to pericentre, where stars are removed  fastest, and the fractional number of escapers at each step is larger in the small-$N$ model because of two effects: (i) the lifetime of the small-$N$ model is smaller (Figs.~\ref{fig:mbound_8k} and \ref{fig:mbound_16k}), while (ii) the time between pericentre passages in the small-$N$ model (which can be inferred from the values of $\Ra$ in Table \ref{tab:data}) is larger.  Therefore, the number of pericentre passages is smaller in the small-$N$ model than in the corresponding large-$N$ model. The rapid mass loss during the pericentre passages implies that dissolution is almost bound to happen around the pericentre, and for this reason the dissolution time of the high-eccentricity models is not really a continuous function of $\epsilon$.  This is important to keep in mind for the forthcoming comparison of lifetimes for different $N$.

We note that the removal of stars at pericentre does not imply that pericentre crossings are the sole mechanism that unbinds stars. For alternative definitions of  bound, for example, being within $\rj$, the $N(T)$ curves display an oscillating behaviour, where $N(T)$  goes up after a pericentre passage (see e.g. Fig.~2 in BM03). This should also not be interpreted as mass gain of the cluster. Both the staircase pattern, and the oscillations, are artefacts of the definition of bound for clusters and illustrate that it is not possible to have a unique definition of the number of bound stars in a cluster on an eccentric orbit.  However, the differences between $N(T)$ for different definitions of bound are small and it is safe to interpret the general trend of $N(T)$ as the evolution of the number of  stars in the cluster. 

Comparing the overall shape of the $N(T)$ curves of the different orbits, we see that there are  similarities.
Core collapse is reached at approximately $T=0.3\tdiss$, after which the escape rate approximately doubles  \citep*{2010MNRAS.409..305L}. For equal-mass models without mass-loss as the result of stellar evolution, the escape rate increases in the pre-collapse phase and this manifests in all curves as a convex curvature (a negative second derivative). After core collapse the escape rate goes as $\dot{N} \propto N^{1/4}$ \citep[equation~\ref{eq:tdisscirc}, and ][]{baumgardt01}, which manifests as a concave curve $N(T)$ (positive second derivative, note that a constant $\dot{N}$ would result in linear $N(T)$ curves). The curvature in pre-collapse and post-collapse evolution is similar for models of different $\epsilon$, though it may be complicated by the `steps' caused by pericentre passages. This trend is not known to apply universally for all Galactic tidal fields, but a discussion of the shapes of $N(t)$ curves is beyond the scope of this paper.

\begin{figure}
	\centering
	\includegraphics[width=8cm]{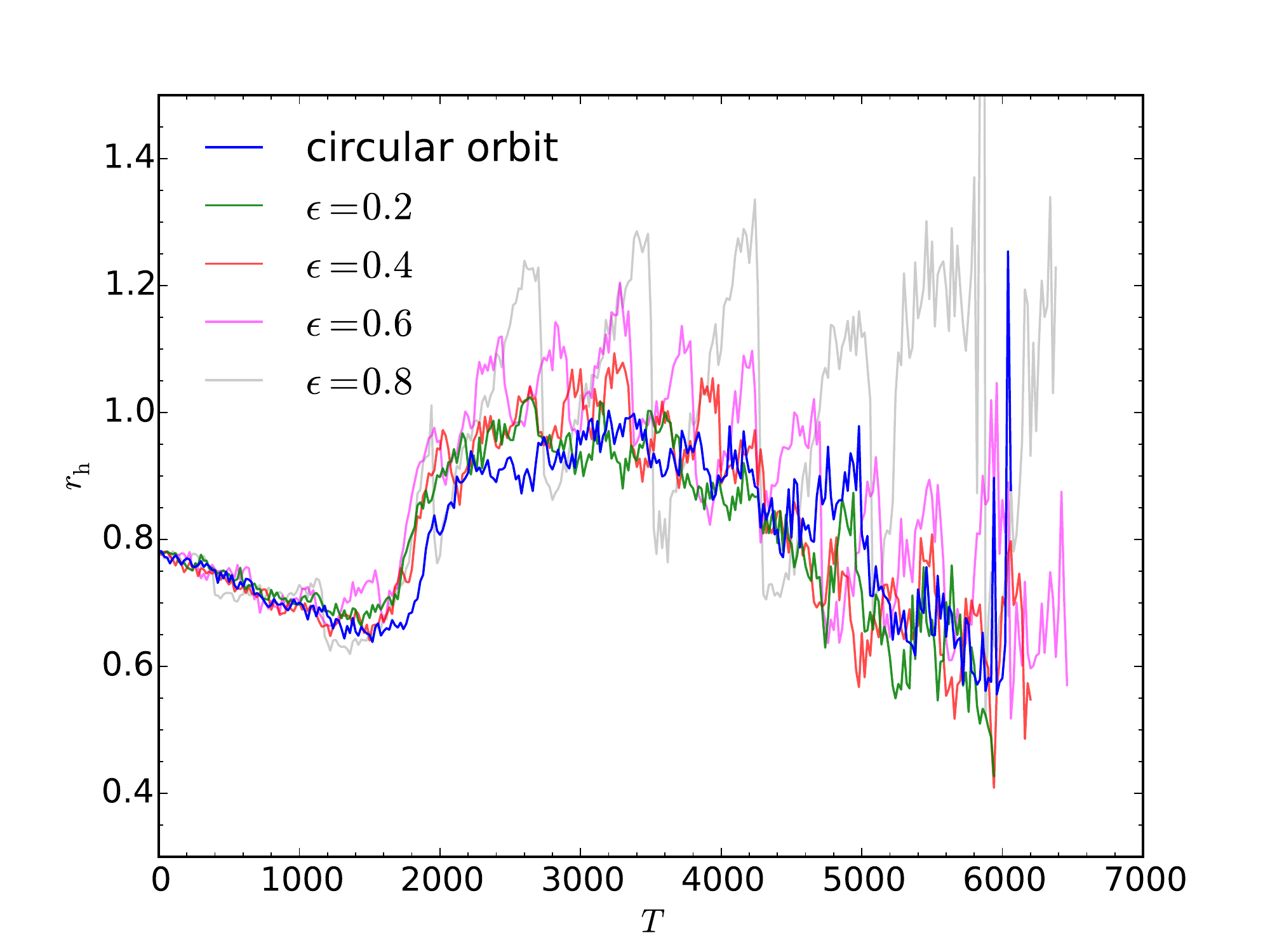} 
	\caption{Half-mass radius $\rh$ evolution of the 8k models with different $\epsilon$ and the same $\tdiss$. }
	\label{fig:rhm_8k}
\end{figure}

\begin{figure}
	\centering
	\includegraphics[width=8cm]{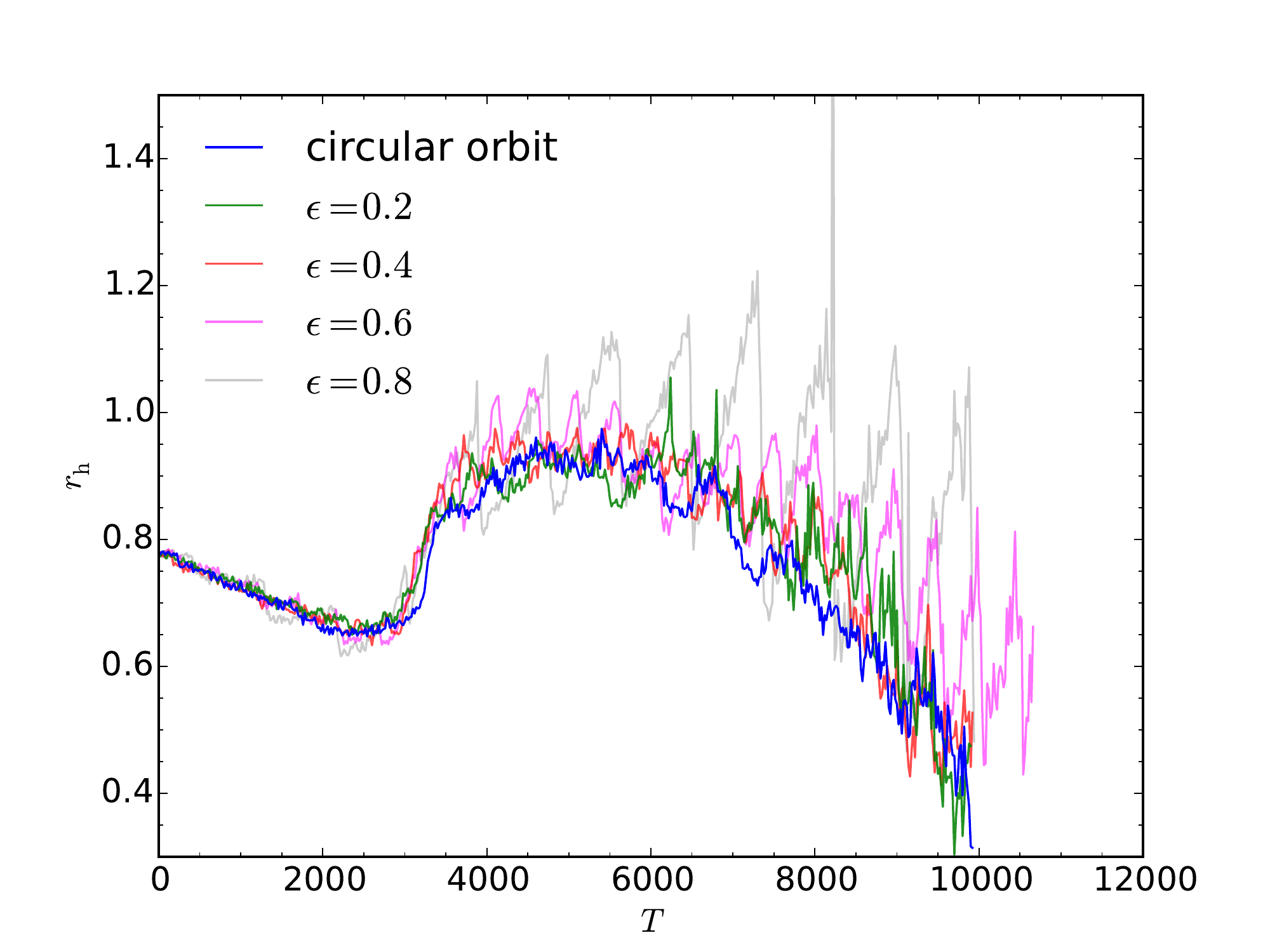} 
	\caption{Same as Fig.~\ref{fig:rhm_8k} but for 16k models.}
	\label{fig:rhm_16k}
\end{figure}

In Fig.~\ref{fig:rhm_8k} and Fig.~\ref{fig:rhm_16k} we show the evolution of $\rh(T)$ of the bound stars for the $N=8\dk$ and $N=16\dk$ models, respectively. As for the $N(T)$, there is general agreement in shape of the $\rh(T)$ curves. All models start with the same initial $\rh\simeq0.78$, and until core collapse,  $\rh$ shrinks as the result of escapers and the absence of a central energy source \citep{2014MNRAS.437..916G}. During the gravothermal catastrophe, $\rh$ increases by about $50\%$, after which it gradually decreases as $N^{1/3}$ \citep{1961AnAp...24..369H}. Similar to the $N(T)$ curves, the $\rh(T)$ curves also exhibit oscillation behaviour and the amplitudes depend on both $N$ and $\epsilon$. During pericentre $\rh$ decreases sharply and then slowly grows until the next pericentre. We note that this behaviour depends on our definition of bound. For example, $\rh$ of all the stars within $\rj$ also oscillates, but has a maximum at $\Ra$.

We recognise similar overall behaviour of $\rh(T)$ in all models, and combined with the similarity between the $N(T)$ curves, we conclude that it is possible to describe the evolution of a cluster on an eccentric orbit, by the evolution of a clusters on a circular orbit with the same $\tdiss$.

\subsection{Scaling of $\tdiss(\Ra, \epsilon)$}
\subsubsection{Results for constant $\tdiss$}
\label{sec:const_tdiss}
\begin{figure}
	\centering
	\includegraphics[width=8cm]{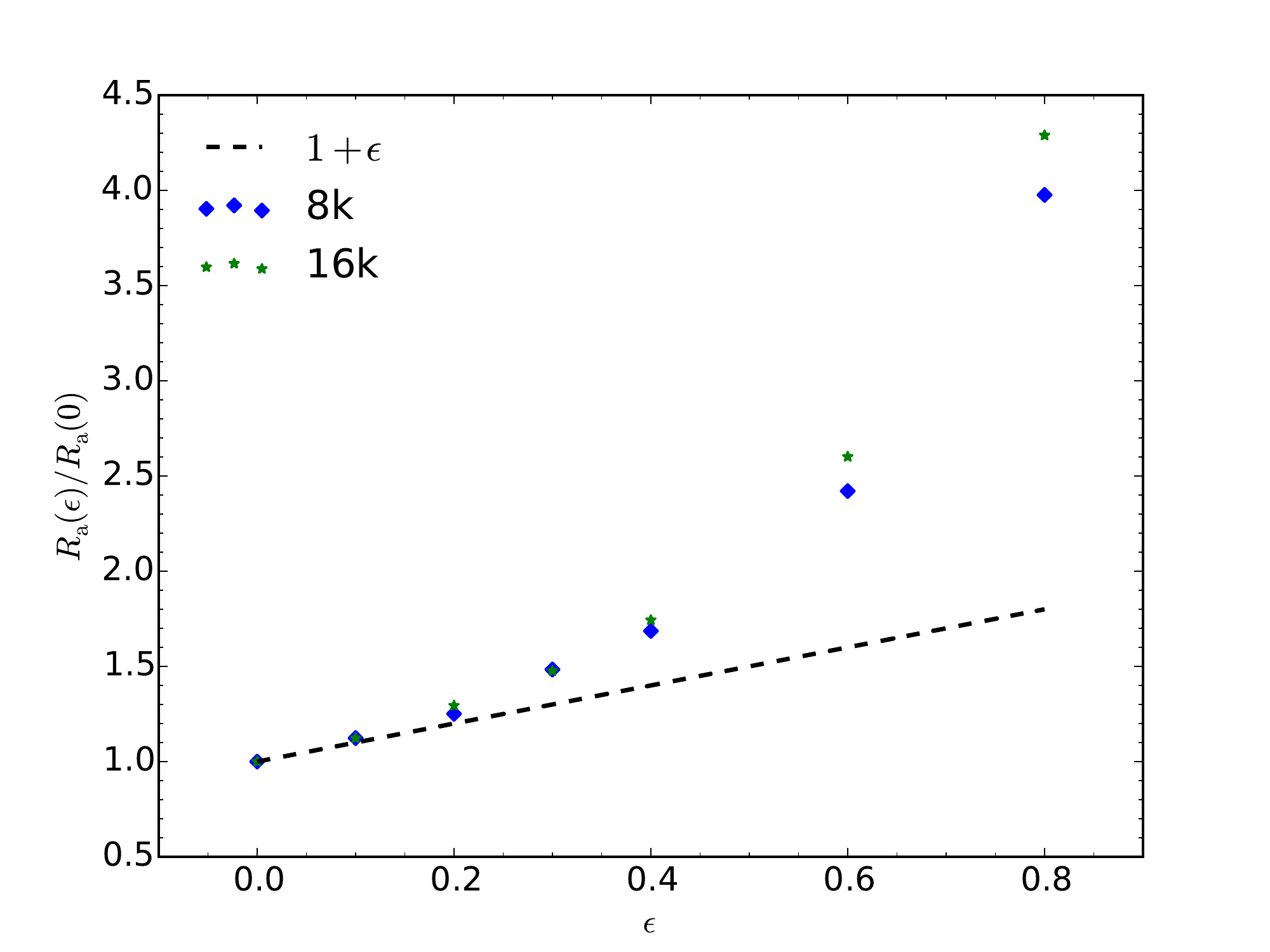} 
	\caption{The apocentre distance $\Ra(\epsilon)$, normalised to $\Ra(0)$ of the circular orbit,   for clusters with the same lifetimes and different eccentricities $\epsilon$ (data from Table~\ref{tab:data}). To first order the data follows the relation $y=1+\epsilon$, corresponding to a constant semi-major axis $a$.}
	\label{fig:ecc_ra}
\end{figure}

In Fig.~\ref{fig:ecc_ra} we show the ratio of $\Ra(\epsilon)$ (Table~\ref{tab:data}) over $\Ra(0)=\RG$ of the circular orbit, for all $\epsilon$ considered. For constant $\Ra$, $\tdiss$ must be a decreasing function of $\epsilon$, and so for increasing $\epsilon$, $\Ra$ must increase to keep $\tdiss$ of the eccentric orbit the same as the circular orbit, and this is indeed what we find. The way $\Ra(\epsilon)$ increases with increasing $\epsilon$ contains information about how $\tdiss$ depends on $\epsilon$.  

In a forthcoming study, Bar-Or et al. (in prep) derive the dependence of $\tdiss$ on $\epsilon$ using perturbation theory. They find that, to first order, $\tdiss$ is
independent of $\epsilon$ for orbits with the same semi-major axis  $a$ (Bar-Or, private communication).  To test this result we plot a line $\Ra(\epsilon) = \Ra(0)(1+\epsilon)$,  corresponding to orbits with the same $a$, because $\Ra(\epsilon) = a(1+\epsilon)$ and for the circular orbit $\RG = a$.  We see that this relation follows the results of our simulations
for $\epsilon\lesssim0.3$ quite well, independently of $N$, and confirming the first order result of Bar-Or~et~al.  But we also consider eccentricities that are much higher than the regime to which the perturbation theory applies.  These empirical results thus serve to quantify the higher order dependence of $\tdiss(a,\epsilon)$ on $\epsilon$, which is the topic of the next sections.

\subsubsection{Results scaled to constant $\Ra$}
\label{sssec:constantRa}
\begin{figure}
	\centering
	\includegraphics[width=8cm]{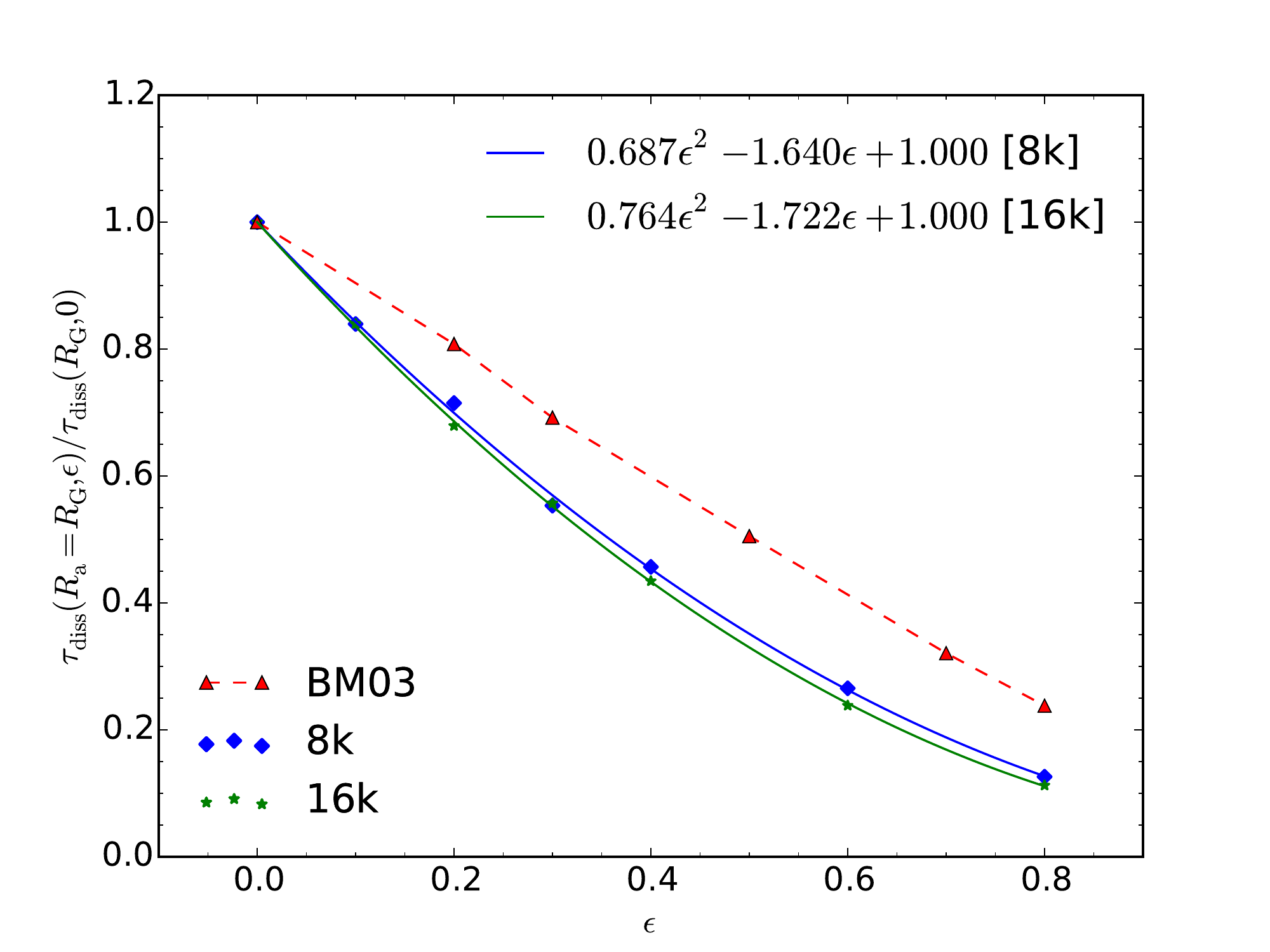} 
	\caption{Dissolution times for all models scaled to the same $\Ra$, normalised to $\tdiss$ of the circular orbit. Solid lines denote polynomial fits to the data (for details, see Sect.~\ref{sssec:constantRa}). Also shown are the results for the $N=32 {\rm k}$ models of BM03. }
	\label{fig:ecc_tdiss_scaled}
\end{figure}

We first want to directly compare our results to the scaling for $\tdiss(\Ra, \epsilon)$ reported by BM03 (equation~\ref{eq:bm03}).
To make the comparison, we need to scale our results such that all our models have the same $\Ra$. We, therefore, need to multiply all our Galactic radii by a scale factor  $R_*=\Ra(0)/\Ra(\epsilon)$. Because for the point-mass Galaxy the radial scale of the cluster is proportional to the Galactic radial scale (equation~\ref{eq:rj}), the scale factor for the cluster's length scale is $r_* = R_*$ and the scale factor for time can be related to the Galactic scale factor as $t_* = R_*^{3/2}$. In Fig.~\ref{fig:ecc_tdiss_scaled} we show the  results for $\tdiss$ scaled to the same $\Ra$, combined with the results of BM03. The $\epsilon$ dependence seems to be stronger in our models, which is suggestive that the mass profile of the Galaxy is important in setting $\tdiss(\epsilon)$. 

However, the difference can be understood (at least for small $\epsilon$) by adopting the hypothesis that $\tdiss$ is independent of $\epsilon$ for fixed $a=\RG$, as in Section \ref{sec:const_tdiss}, i.e. that $\tdiss(\Ra = \RG(1+\epsilon),\epsilon) = \tdiss(\RG,0)$.  Since $t_* = R_*^{3/2}$ for the point-mass galaxy, we deduce that $\tdiss(\Ra=\RG,\epsilon) = \tdiss(\RG,0)/(1+\epsilon)^{3/2} \simeq \tdiss(\RG,0)(1 - 1.5\epsilon)$.  For the singular isothermal model of BM03, however, $t_* = R_*$, and so the corresponding result is $\tdiss(\Ra=\RG,\epsilon) \simeq \tdiss(\RG,0)(1 -\epsilon)$.  Our hypothesis therefore explains why the dependence on $\epsilon$ in Fig.~\ref{fig:ecc_tdiss_scaled} is steeper for our models than in BM03, for small $\epsilon$, and unifies the results of the two studies in this regime.

Fig.~\ref{fig:ecc_tdiss_scaled} also gives second-order polynomial fits\footnote{It could be argued that the lifetime should be zero for $\epsilon = 1$, and the quadratic fits provided in Fig.~\ref{fig:ecc_tdiss_scaled} are inconsistent with this, but those in Fig.~\ref{fig:compare_bm03} accommodate this idea. On the other hand, the lifetime of the model can hardly be less than the
time taken to reach perigalacticon.} to our results, and the foregoing argument approximately explains the first-order coefficient of $\epsilon$ in these fits. BM03 themselves showed that the factor $1-\epsilon$ gave a satisfactory fit to their models for the entire range of $\epsilon$.  By reversing the above argument we easily see that this result is consistent with a hypothesis that $\tdiss(\Ra = \RG(1+\epsilon),\epsilon) = \tdiss(\RG,0)(1-\epsilon^2)$, which we discuss further in the next section, where we scale all results to orbits with the same $a$.

\subsubsection{Results scaled to constant semi-major axis $a$}
\label{sssec:constant_a}
To be able to compare all results to the theoretical prediction by Bar-Or et al., we  present all results scaled to orbits with the same $a$. We note that the results of this exercise for the BM03 models should be interpreted with caution, because their models include the effect of stellar evolution, which imposes a fixed physical time-scale. Bearing this word of caution in mind, we scale the Galactic orbits with $R_*= (1+\epsilon)\Ra(0)/\Ra(\epsilon)$. The radial scale factor of the cluster itself is dependent on the mass profile: for the point-mass galaxy $r_* = R_*$ as before, while for the singular isothermal halo $r_* = R_*^{2/3}$. The scale factors for time for the two Galaxy mass models are related to $R_*$ as $t_* = R_*^{3/2}$ and $t_* = R_*$, respectively. 

\begin{figure}
	\centering
	\includegraphics[width=8cm]{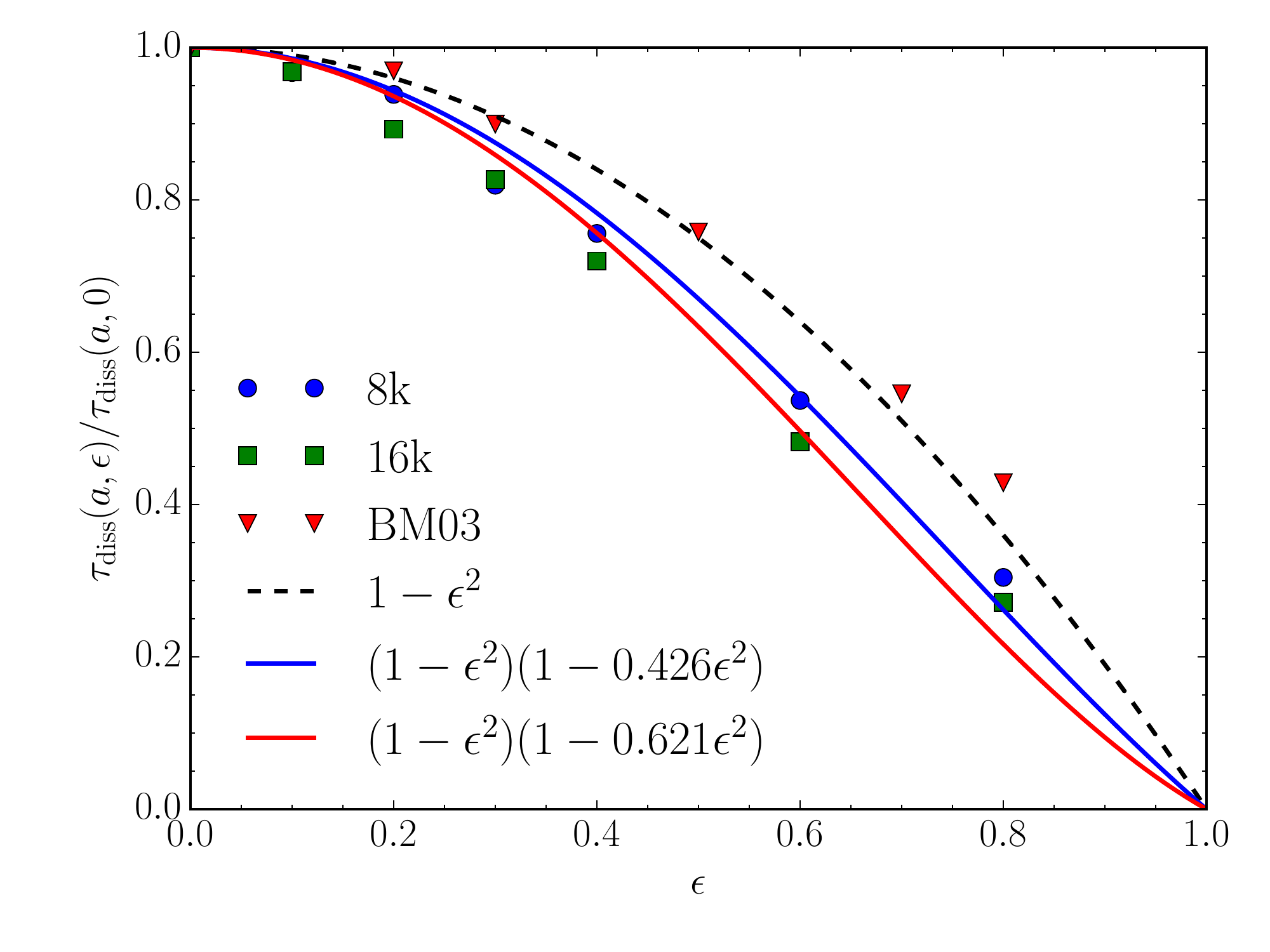} 
	\caption{Dissolution time $\tdiss$ for different $\epsilon$, normalised to the result for the circular orbit, for the models discussed in this paper and BM03 (32k models). All results have been scaled to orbits with the same semi-major axis $a$. Simple even polynomial functions, up to fourth order in $\epsilon$, are shown for comparison (for details, see Sect. \ref{sssec:constant_a}).}
	\label{fig:compare_bm03}
\end{figure}

In Fig.~\ref{fig:compare_bm03} we present the results of all models, scaled to the same $a$ and normalised to the circular orbit.  
We note that \citet{webb14} studied eccentric orbits with direct $N$-body models with similar properties as BM03, and they compared a model with high eccentricity ($\epsilon=0.9$) to a model on a circular orbit with approximately the same lifetime. Applying the same scaling to their values of $\Ra$ we find that their $\epsilon=0.9$ data point would extend the trend of the BM03 scaling. Our results  are slightly below the results of BM03. It is tempting to explain this offset to the difference in Galactic mass model: the point-mass model has stronger tidal forces at peri-centre and, therefore, one way of interpreting the difference in Fig.~\ref{fig:compare_bm03} is that clusters on eccentric orbits dissolve faster in such halos.

However, the small difference can perhaps also be explained by differences in how the clusters were setup relative to the tides and the differences in treatment of stellar evolution. BM03 consider King models with Roche-filling initial conditions for the models on circular orbits, which means that the truncation radius of the King model equals $\rj$. For their models on circular orbits the initial ratio $\rh/\rj \simeq 0.19$ (Fig.~\ref{fig:rh_rj}), which is somewhat larger than what is adopted here ($\rh/\rj=0.1$).  
 
In addition, their models consider stellar evolution mass loss, and for the clusters on circular orbits, a fraction of the stars is pushed over the tidal boundary as the result of the expansion due to stellar mass loss \citep{2010MNRAS.409..305L},  shortening the lifetimes of the models on circular orbits by a mechanism that is not included in our models \citep*[for a discussion on the sensitivity of the Roche-filling models of BM03 to stellar mass loss, see][]{2015MNRAS.449L.100C}. 

For eccentric orbits, BM03 fix $\rh$ to the value a cluster would have on a circular orbit at  $\Rp$ of the eccentric orbit. During the evolution, $\rj$ is time-dependent, but motivated by our  earlier finding, we can define $\rj$ for the eccentric orbit as the Jacobi radius of an identical cluster on a circular obit with the same $\tdiss$. Using that definition, the BM03 models initially have $\rh/\rj \propto  (1+\epsilon)^{-2/3}$, whereas in our models $\rh/\rj = 0.1$, for all $\epsilon$. Therefore,  the BM03 models with high $\epsilon$ are more compact with respect to the tides than our models, which could result in slightly larger $\tdiss$ compared to our models at the same $\epsilon$. We are therefore cautious with interpreting the small difference between the BM03 results, and ours, as being due to difference in Galactic mass profile. 

In order to quantify the higher order dependence of $\tdiss$ on $\epsilon$, we  plot two simple functions in Fig.~\ref{fig:compare_bm03}. The functional form $y=1-\epsilon^2$ is motivated by the results of BM03, because 
this relation is what follows when scaling the result reported by BM03 (equation~\ref{eq:bm03}) to orbits with constant $a$ (Section~\ref{sssec:constantRa}). As expected, this relation describes the BM03 results very well. However, it over-predicts $\tdiss$ of our models, with a maximum difference of about $20\%$ at $\epsilon = 0.6$. 

Motivated by our empirical findings, and the theoretical work of Bar-Or et al., we speculatively propose a characterisation of the higher order dependence of $\tdiss$ on $\epsilon$, on the basis of a simple symmetry argument. Indeed, it can be argued that the quantity $\tdiss(a,\epsilon)/\tdiss(a,0)$ should naturally be an even function of $\epsilon$, which implies that, in a series expansion in $\epsilon$, the odd terms of any order must vanish. This expectation on the parity of the function follows if we assume that the lifetime is independent of the initial phase on the Galactic orbit, as reversing the sign of $\epsilon$ simply corresponds to starting at perigalacticon instead of (as in our models) at apogalacticon.  We therefore consider an even,  polynomial function such that $y(\epsilon)=0$ for $\epsilon =1$: $y(x) = (1-\epsilon^2)(1+(c+1)\epsilon^2)$, in which the constant term is imposed to be 1 by virtue of the chosen normalisation $\tdiss(a,\epsilon)/\tdiss(a,0)$.  We have then determined the best-fit value of the free coefficient $c$ for the two sets of models with $N=8\rm{k}$ and $N=16\rm{k}$, as depicted in Fig. \ref{fig:compare_bm03}.

In consideration of the simplicity of our argument, based on a reasonable but unproven assumption, we encourage the reader to accept the values resulting from the fitting process only for empirical guidance, as the full perturbation analysis of the escape problem, which is needed to constrain analytically the coefficient $c$, is beyond the scope of this article (see Bar-Or et al., in preparation).

In addition, given the relatively low $N$ of our models compared to the BM03 simulations, and other differences in the initial conditions between the two sets of models (see the discussion above on differences in the initial $\rh/\rj$), we are cautious with concluding that these different scalings as being due to the different Galactic mass models.

These results serve as benchmarks for future theoretical work on the dissolution of clusters on eccentric orbits in different Galactic potentials.

\section{Conclusions}
\label{sec:conclusions}
We model star clusters on circular and eccentric orbits with direct $N$-body simulations in order to gain insight in  the evolution of cluster properties at different eccentricities $\epsilon$, and the scaling relations for the dissolution timescale ($\tdiss$) as a function of $\epsilon$. We deploy direct $N$-body simulations to model idealised systems of $N=8\dk$ and $N=16\dk$ stars of the same mass, on orbits around a point-mass galaxy. For the models on eccentric orbits, we iteratively find the apogalactic radius $\Ra$ on which the cluster $\tdiss$ is the same as for the circular orbit.

When scaling our results to orbits with the same semi-major axis $a$, we find that $\tdiss$ is, to first order,  independent of $\epsilon$. We show that this scaling agrees with results presented by BM03, who modelled clusters with a mass spectrum and the effects of stellar mass loss in singular isothermal Galactic halos. Their results suggest slightly longer $\tdiss$ at higher $\epsilon$ than found here, which can be explained by differences in the initial $\rh$ with respect to the tidal truncation. Alternatively, there may be a dependence on Galactic mass profile, in the sense that $\tdiss$ is more sensitive to $\epsilon$  in the case of point-mass galaxies. This explanation has  theoretical support, because the heating at perigalactic passages in a point-mass model is more severe than in the extended singular isothermal model \citep{1999ApJ...514..109G}. Because of the many differences between our models and the BM03 we are cautious with interpreting the small difference between our results and BM03 in either direction.

Finally, we quantify the higher order dependence of $\tdiss(a, \epsilon)$ on $\epsilon$. A relation of the form $\tdiss(a, \epsilon) = f(\epsilon)\tdiss(a, 0)$, with $f(\epsilon) = 1-\epsilon^2$ describes the results of BM03 very well. For the models presented here, a functional form of $f(\epsilon) = (1-\epsilon^2)(1-c\epsilon^2)$, with $c\simeq0.5$, is more accurate. These data serve as benchmark for future theoretical work on the dissolution of clusters on eccentric orbits. 

We find that clusters with the same initial $N$, $r_{\rm h}$ and $\tdiss$, but different $\epsilon$, have similar evolution of the number of bound stars and half-mass radius $\rh$. This implies that we can approximate the evolution of properties of clusters on eccentric orbits by that of clusters on circular orbits. This is useful  for modelling techniques  that are not able to include orbital eccentricity, such as the Fokker-Planck method, or the Monte Carlo method and/or time-dependent Galactic tides. For example, \citet{2008MNRAS.389.1858H} present Monte Carlo models of the Galactic globular cluster M4, which is on an eccentric orbit. The authors model M4 on a circular orbit, with approximately the same $\tdiss$ as M4 has on its eccentric orbit. Our results confirm that  this approach is valid and  results in a representative evolution of $N$ and $\rh$ in these models. Another application of our result can be found in semi-analytic models of cluster evolution \citep[e.g.][]{2014ApJ...785...71G}. 
The fast cluster evolution code Evolve Me A Cluster of StarS \citep[$\emacss$,][]{2012MNRAS.422.3415A,2014MNRAS.437..916G, 2014MNRAS.442.1265A}  solves a set of coupled differential equations for the rate of change of $\rh$, $\rj$ and $N$. The method requires an expression for $\dot{N}$ that depends on the tidal field. The results in this study can be used to include orbital eccentricity in \emacss\ by using the functional form for $\tdiss(a,\epsilon)$ to include $\epsilon$ in the $\dot{N}$ term.

Finally, several  models for generating tidal tails of globular clusters have recently been developed
\citep{2012MNRAS.420.2700K, 2014ApJ...795...95B,2015MNRAS.450..575A}. These models require as input an escape rate of stars from the cluster. Our analytic expression for $\tdiss(a,\epsilon)$ can be used to obtain expressions for the average escape rate from clusters on eccentric orbits.

\section{Acknowledgements}
We wish to thank the anonymous referee for her/his comments that helped to improve the manuscript considerably. This work was initiated during the International Summer-Institute for Modeling in Astrophysics (ISIMA) in 2014, hosted at CITA at the University of Toronto. We gratefully thank the organizers of ISIMA, especially Pascale Garaud, for the kind support of our participation. We thank CITA for hosting the ISIMA program. We thank Florent Renaud, Rainer Spurzem and Thijs Kouwenhoven for useful discussions. MXC acknowledges supported by Chinese Academy of Sciences Grant Number 2009S1-5, and
through the ``Thousand Talents" (Qianren) program of China (R. Spurzem). MG acknowledges financial support from the Royal Society (University Research Fellowship) and the European Research Council (ERC-StG-335936, CLUSTERS), and ALV from the Royal Commission for the Exhibition of 1851. All authors are grateful to Keigo Nitadori and Sverre Aarseth for making {\tt NBODY6} and its GPU enabled version publicly available and to Dave Munro of the University of Surrey for his support of the GPU cluster at the University of Surrey.

\bibliographystyle{mn2e}
\bibliography{total.bib}
\bsp


\end{document}